\documentclass[12pt]{article}
\usepackage{epsfig}
\newcommand{\beq}{\begin{equation}}
\newcommand{\eeq}{\end{equation}}
\newcommand{\beqa}{\begin{eqnarray}}
\newcommand{\eeqa}{\end{eqnarray}}
\newcommand{\ba}{\begin{array}}
\newcommand{\ea}{\end{array}}

\begin{document}

\title{Condensate fraction in metallic superconductors 
and ultracold atomic vapors} 
\author{Luca Salasnich} 

\date{\small{CNR and CNISM, Dipartimento di Fisica ``Galileo Galilei'', \\
Universit\`a di Padova, Via Marzolo 8, 35131 Padova, Italy \\
e-mail: luca.salasnich@cnr.it}}

\maketitle

\begin{abstract}
We investigate the condensate density and the condensate 
fraction of conduction electrons 
in weak-coupling superconductors by using the BCS 
theory and the concept of off-diagonal-long-range-order. 
We discuss the analytical formula of the zero-temperature condensate 
density of Cooper pairs as a function of Debye frequency and 
energy gap, and calculate the condensate fraction 
for some metals. We study the density of Cooper pairs 
also at finite temperature showing its connection with 
the gap order parameter and the effects of the electron-phonon coupling. 
Finally, we analyze similarities and differences 
between superconductors and ultracold Fermi atoms in the determination 
of their condensate density by using the BCS theory.
\end{abstract}


\section{Introduction}

The condensate fraction of fermionic alkali-metal atoms 
has been recently investigated 
\cite{sala-odlro,ortiz,ohashi,sala-odlro2} by using extended 
BCS (EBCS) equations 
\cite{eagles,leggett,noziers,sademelo,engelbrecht,marini}  
from the BCS regime of Cooper-pairs to the BEC regime of molecular dimers 
\cite{sala-odlro,ortiz,ohashi}. In particular, we have found 
\cite{sala-odlro} a remarkable agreement between this simple 
mean-field theory and the experimental results 
\cite{zwierlein,zwierlein2}. These results  
indicate the presence of a relevant fraction of condensed pairs of $^6$Li 
atoms also on the BCS side of the Feshbach resonance. 
Monte Carlo calculations \cite{astrakharchik}  
have shown that the zero-temperature 
mean-field predictions \cite{sala-odlro,ortiz} slightly 
overestimate the condensed fraction of Fermi pairs. 
Very recently it has been reported \cite{inada} an 
accurate measurement of the temperature dependence 
of the condensate fraction 
for a fermion pair condensate of $^{6}$Li atoms 
near the unitarity limit of the BCS-BEC crossover. 
Also these new experimental data \cite{inada}
are in agreement with mean-field 
theoretical predictions at finite temperature \cite{ohashi}. 
\par 
In superfluids made of ultracold atoms, 
the inter-atomic interaction is attractive for all fermions 
of the system \cite{leggett}. On the contrary, 
in metallic superconductors there is an attractive interaction 
between fermions only near the Fermi surface \cite{leggett1,fetter}. 
As a consequence, the condensate fraction of metallic superconductors 
has distinctive properties with respect to those of atomic superfluids. 
Despite the BCS theory is $52$ years old \cite{bardeen}, 
the condensate fraction of Cooper pairs 
in superconductors has been considered only in few papers 
\cite{yang,dunne,campbell,sudip} and in the recent book 
of Leggett \cite{leggett1}. In fact, 
in superconductors the condensate fraction has never been 
measured: only very recently Chakravarty and Kee have proposed 
to measure it by using magnetic neutron scattering \cite{sudip}. 
\par 
In this paper we analyze in detail the condensate density of 
conduction electrons in weak-coupling superconductors 
at zero and finite temperature 
by using BCS theory \cite{bardeen} and the concept of 
off-diagonal-long-range-order \cite{yang,penrose}. 
For the first time, we calculate explicitly the density of electronic 
Cooper pairs and the condensate fraction 
for various metals and show its dependence 
on the Debye frequency, the electron-phonon interaction and 
the energy gap. Another novelty of this paper is the analytical and 
numerical investigation of the temperature 
dependence of the condensate fraction, for which we find a power-law behavior. 
Finally, we compare of the BEC theory of 
superconductors with the extended BEC theory of ultracold Fermi atoms 
for obtaining the condensate density and the condensate fraction. 

\section{BCS theory and ODLRO}

The BCS Lagrangian density of conduction electrons 
with spin $\sigma={\uparrow},{\downarrow}$ 
near the Fermi surface is given by 
\beq 
{\hat {\cal L}} = \sum_{\sigma} {\hat \psi}_{\sigma}^+ 
\left( i\hbar {\partial \over \partial t} 
-\epsilon({\nabla}) + \mu \right) { {\hat \psi}}_{\sigma} + g \; 
{\hat \psi}^+_{\uparrow}
{\hat \psi}^+_{\downarrow} {\hat \psi}_{\downarrow} 
{\hat \psi}_{\uparrow} \; ,
\label{lagr}
\eeq
where ${\hat \psi}_{\sigma}({\bf r},t)$ is the electronic 
field operator which satisfies the familiar equal-time 
anti-commutation rules of fermions. 
Here $\epsilon({\nabla})$ is the differential operator such that 
$\epsilon({\nabla}) e^{i{\bf k}\cdot{\bf r}} = \epsilon_k 
e^{i{\bf k}\cdot{\bf r}}$, where $\epsilon_k$ is the energy 
spectrum of conduction electrons in a specific metal \cite{umezawa}. 
The attractive interaction between electrons is described by a contact 
pseudo-potential of strength $g$ ($g>0$). For metals 
this electron-phonon interaction strength is attractive 
only for conduction electrons near the Fermi surface 
\cite{leggett1,fetter,bardeen}. The chemical potential $\mu$ 
fixes the number $N$ of conduction electrons. 
 
The Heisenberg equation of motion of the field operator 
${\hat \psi}_{\uparrow}({\bf r},t)$ can be immediately 
derived and reads 
\beq 
i \hbar {\partial\over \partial t} {\hat \psi}_{\uparrow} = 
\left[ \epsilon({\nabla}) - \mu \right] { {\hat \psi}}_{\uparrow} 
- g \; {\hat \psi}^+_{\downarrow} {\hat \psi}_{\downarrow} 
{\hat \psi}_{\uparrow} \; .  
\label{eom} 
\eeq 
In the BCS theory the interaction term of Eq. (\ref{eom}) can be treated 
within the minimal mean-field approximation 
$
{\hat \psi}^+_{\downarrow}
{\hat \psi}_{\downarrow}
{\hat \psi}_{\uparrow} 
\simeq 
{\hat \psi}^+_{\downarrow} 
\langle 
{\hat \psi}_{\downarrow}
{\hat \psi}_{\uparrow} 
\rangle  
$. 
In this way Eq. (\ref{eom}) becomes 
\beq 
i \hbar {\partial\over \partial t} {\hat \psi}_{\uparrow} = 
\left[ \epsilon({\nabla}) - \mu \right] 
{\hat \psi}_{\uparrow} - \Delta \ {\hat \psi}^+_{\downarrow} 
\; , 
\label{eom1} 
\eeq 
where 
\beq 
\Delta({\bf r},t) = g \ \langle 
{\hat \psi}_{\downarrow}({\bf r},t) \, 
{\hat \psi}_{\uparrow}({\bf r},t) 
\rangle
\label{delta}
\eeq
is the gap function. The condensate wave function of Cooper pairs 
\cite{leggett1,campbell} is instead given by 
\beq 
\Xi({\bf r},{\bf r}',t) = \langle 
{\hat \psi}_{\downarrow}({\bf r},t)\; 
{\hat \psi}_{\uparrow}({\bf r}',t) \rangle \; .
\eeq
As shown by Yang \cite{yang}, this two-particle wave function is 
strictly related to the largest eigenvalue $N_0$ of 
two-body density matrix of the system. 
$N_0$ gives the number of Fermi pairs 
in the lowest state, i.e. the condensate 
number of Fermi pairs \cite{leggett1,campbell,yang}, and it can be written as 
\beq
N_0 = \int |\Xi({\bf r},{\bf r}',t)|^2 \ d^3{\bf r} \, d^3{\bf r}' \; . 
\label{number0}
\eeq 
A finite value for the condensate fraction $f=N_0/(N/2)$ 
in the thermodynamic limit $N\to\infty$ implies 
off-diagonal-long-range-order \cite{yang,penrose}. 

\section{Gap equation and condensate density}

To investigate the properties of 
the condensate fraction of electronic pairs 
we adopt the following Bogoliubov representation of the field operator 
\beq 
{\hat \psi}_{\uparrow}({\bf r},t)= \sum_{\bf k} 
\left( {u_k\over V^{1/2}} e^{i({\bf k}\cdot {\bf r} 
- \omega_k t)} {\hat b}_{{\bf k}\uparrow} 
- {v_k\over V^{1/2}} 
e^{-i({\bf k}\cdot {\bf r} -\omega_k t)} 
{\hat b}_{{\bf k}\downarrow}^+ \right) 
\label{field}
\eeq 
in terms of the anti-commuting quasi-particle 
Bogoliubov operators ${\hat b}_{{\bf k}\sigma}$, 
with $V$ the volume of the system and $E_k=\hbar \omega_k$ 
the excitation energies of quasi-particles \cite{leggett1,fetter}. 

The thermal averages of quasi-particle Bogoliubov 
operators are given by 
\beq
\langle 
{\hat b}_{{\bf k}\sigma}^+ 
{\hat b}_{{\bf k}'\sigma'} 
\rangle = {1 \over e^{\beta E_k} + 1} \ 
\delta_{{\bf k} {\bf k}'} \delta_{\sigma \sigma'}
= \bar{n}_k \ \delta_{{\bf k} {\bf k}'} \delta_{\sigma \sigma'} \; , 
\eeq
where $\beta=1/(k_B T)$ with $k_B$ the Boltzmann constant,  
$T$ the absolute temperature, and $\bar{n}_k$ is the thermal 
Fermi distribution. 

By using these results, the gap function, Eq. (\ref{delta}), becomes 
\beq 
\Delta = \frac{g}{V}{ {\sum_{\bf k}}' } (1-2 \bar{n}_k) u_k v_k \; ,   
\label{gap}
\eeq
while the condensate number of conduction electrons, Eq. (\ref{number0}), 
satisfies this expression \cite{sala-odlro,campbell}
\beq 
N_0 ={ {\sum_{\bf k}}' } (1-2 \bar{n}_k)^2 u_k^2 v_k^2 \; .   
\label{cond}
\eeq 

The 'prime' restricts the summation to states within a shell of width 
$\hbar \omega_D$ about the Fermi surface.

To determine the amplitudes $u_k$ and $v_k$ of quasi-particles, 
one inserts Eq. (\ref{field}) into Eq. (\ref{eom1}) and 
obtains the familiar Bogoliubov-de Gennes equations, which give 
\beq 
u_k^2 = {1\over 2} \left( 1 + \frac{\xi_k}{E_k} \right) \, , \quad 
v_k^2 = {1\over 2} \left( 1 - \frac{\xi_k}{E_k} \right) \, ,    
\eeq 
where 
\beq 
\xi_k = \epsilon_k - \mu \; , \quad 
E_k=\sqrt{\xi_k^2 + \Delta^2} \; . 
\eeq
Eqs. (\ref{gap}) and (\ref{cond}) can then be written as 
\beq
\Delta =\frac{g}{V}{ {\sum_{\bf k}}' } {\Delta \over 2E_k} 
\tanh({\beta E_k\over 2})   
\label{gap-0}
\eeq
\beq
N_0 ={ {\sum_{\bf k}}' } {\Delta^2 \over 4E_k^2} 
\tanh^2({\beta E_k\over 2})   
\label{cond-0} 
\eeq
where $\tanh(\beta E_k/2)  = 1-2 \bar{n}_k$. 

In the thermodynamic limit, where the volume $V$ goes to infinity, 
$\sum_{\bf k}$ can be replaced by 
$V\int d^3{\bf k}/(2\pi)^3=V\int N(\xi) d\xi$ 
with $N(\xi)=\int d^3{\bf k}/(2\pi)^3 \ \delta(\xi -\xi_k)$. 
In metals the condition $\hbar\omega_D\ll \mu$ is always 
satisfied \cite{fetter}, consequently we can use the approximation 
$\int N(\xi) d\xi \simeq N(0) \int d\xi$, where 
\beq 
N(0)= \int {d^3{\bf k}\over (2\pi)^3} \ \delta(\mu - \epsilon_{\bf k}) 
\eeq
is the density of states at the Fermi surface.  
In this way the previous equations (\ref{gap-0}) and (\ref{cond-0}) become 
\beq
{1\over gN(0)} = \int_0^{\hbar\omega_D} 
{ \tanh({\beta \over 2} \sqrt{\xi^2+\Delta^2}) 
\over \sqrt{\xi^2+\Delta^2} } \ d\xi 
\label{gap-1}
\eeq
\beq 
n_0 = {1\over 2} N(0) \Delta^2 \int_{0}^{\hbar\omega_D} 
{ \tanh^2({\beta \over 2} \sqrt{\xi^2+ \Delta^2}) 
\over \xi^2+\Delta^2} \ d\xi 
\label{cond-1}
\eeq
where $n_0=N_0/V$ is the density of electrons in the condensate. 

\subsection{Zero-temperature condensate}

Let us consider first the zero-temperature case ($T=0$). 
From Eqs. (\ref{gap-1}) and (\ref{cond-1}) we get the zero temperature 
energy gap $ \Delta (0)$:
\beq 
{1\over g N(0)} 
= \ln\left( {\hbar\omega_D\over \Delta(0)} 
+ \sqrt{1+{\hbar^2\omega_D^2\over \Delta(0)^2}}\right) 
\; , 
\label{rigap}
\eeq
and the zero-temperature condensate density:
\beq 
n_0(0) = {1\over 2} N(0) \Delta(0) \arctan({\hbar\omega_D \over \Delta(0)}) 
\; . 
\label{cond-den}
\eeq
This expression shows that the condensate density $n(0)$ can be 
expressed in terms of density of states $N(0)$, 
energy gap $\Delta(0)$ and Debye energy $\hbar\omega_D$. 
Finally, the zero-temperature condensate fraction 
$f(0)=n_0(0)/(n/2)$ is given by 
\beq 
f(0) = {1 \over 2} {N(0)\over n} \Delta(0) 
\arctan({\hbar\omega_D \over \Delta(0)}) 
\; ,    
\label{fraction}
\eeq 
where $n$ is the density of conduction electrons.

Under the condition $\Delta(0) \ll \hbar\omega_D$, 
from Eqs. (\ref{rigap}) we find the familiar weak-coupling BCS result 
\beq 
\Delta(0) = 2 \hbar\omega_D \, \exp(-{1\over gN(0)}) \;  
\label{e1}
\eeq
for the energy-gap order parameter, 
while the condensate density (\ref{cond-den}) can be written 
as \cite{sudip,leggett1}:
\beq 
n_0(0) = {\pi \over 4} N(0) \Delta(0)  
\label{e2}
\eeq

We stress that in many real superconductors 
 the simple BCS theory reported above 
is not accurate, and one has to take into account the retarded 
electron-electron interaction via phonons \cite{eliashberg} 
and also the Coulomb repulsion \cite{mcmillan}. 
The results obtained above by using the mean-field BCS theory 
are reliable only in the weak-coupling regime, 
i.e. for $\Delta(0) \ll \hbar\omega_D$, where $gN(0) \leq 0.3$. 
Therefore we will continue our analysis of the
condensate fraction only for a class of superconductors which satisfy 
this condition. 

Within the free-electrons 
Sommerfeld approximation, where the energy spectrum 
$\epsilon_k$ of conduction electrons has the simple quadratic 
behavior $\epsilon_k=\hbar^2k^2/(2m^*)$, the free particle 
density of states $N_{free}(0)$ is related to the total density 
of conduction electrons by $n=4N_{free}(0)\mu/3$ and the zero-temperature 
condensate fraction reads $f(0)=3\pi \Delta(0)/(8\mu)$. 
To get a better estimate, we correct the free electron value of 
$N(0)$ by an effective mass as obtained from specific heat measurements, 
i.e. we use the expression $N(0)= (m^*/m) N_{free}(0)$. 

\begin{table}
\begin{center}
\begin{tabular}{|c|c|c|c|c|}
\hline
~~~~ & $n_0(0)$ [$10^{-33}$ m$^{-3}$] & 
$f(0)$ [$10^{-5}$] & $g N(0)$ & $T_c$ [K] \\
\hline
Cd & $4.18$ & $0.9$ & $0.179$ & $0.51$
\\
Zn & $7.72$ & $1.2$ & $0.172$ & $0.79$
\\
Al & $22.4$ & $2.5$ & $0.168$ & $1.15$
\\
Tl & $31.0$ & $5.9$ & $0.263$ & $2.43$
\\
In & $62.1$ & $10.8$ & $0.267$ & $3.46$
\\
Sn & $62.5$ & $8.4$ & $0.254$ & $3.68$
\\
\hline
\end{tabular}
\end{center}
\vskip 0.3cm
\caption{BCS predictions for weak-coupling 
superconducting metals: $n_0$ is the zero-temperature 
condensate density, obtained with Eq. (\ref{e2}); 
$f(0)=n_0(0)/(n/2)$ is the zero-temperature condensate fraction; 
$gN(0)$ is the electron-phonon strength, 
calculated with Eq. (\ref{rigap}). $T_c$ is the critical 
temperature from  Eq. (\ref{TC}).}
\end{table}

In the first two columns of Tab. 1 we show the zero-temperature 
condensate density $n_0$ and condensate fraction $f(0)$ 
of simple metals obtained from Eqs. (\ref{cond-den}) and (\ref{fraction}) 
by using the experimental data of $\Delta(0)$ and $\omega_D$ 
obtained from Ref. 24 (when the comparison is possible, they agree 
within a few percent with those reported in Ref. 25).
In the third column we report the electron-phonon strength 
$gN(0)$ calculated with Eq. (\ref{rigap}) 
knowing $f(0)$. The table shows that indeed these simple metals are all 
in the weak-coupling regime. For completeness, in the forth column 
we insert the theoretical determination (see Eq. (\ref{TC})) 
of the critical temperature $T_c$, which is very reliable 
for these simple metals, when compared with the experimental data. 

\subsection{Finite-temperature condensate} 

Let us now investigate the behavior of the condensate density $n_0$ 
at finite temperature $T$. Under the condition $\hbar\omega_D \gg k_B T_c$, 
which is always satisfied, near $T_c$ the energy gap goes to zero 
according to the power law \cite{fetter,leggett1} 
\beq 
\Delta(T) = 3.06 \ k_B T_c \ \left(1 - {T\over T_c} \right)^{1/2} \; . 
\eeq
Instead for the condensate density $n_0(T)$, from 
Eq. (\ref{cond-1}) and the previous expression, we find near $T_c$
\beq 
n_0(T) = 0.43 \, N(0) {\Delta(T)^2\over k_B T_c} = 
4.03 \, N(0) {k_B T_c} \ \left(1 - {T\over T_c} \right) \; .  
\eeq

For a generic temperature $T$ we solve numerically 
Eqs. (\ref{gap-1}) and (\ref{cond-1}). 
The theoretical critical temperature $T_c$, obtained 
from Eq. (\ref{gap-1}) setting $\Delta(T_c)=0$, 
is given by the well-known result \cite{fetter}
\beq 
k_B T_c = 1.13 \ \hbar \omega_D \ \exp(-{1\over gN(0)}) \; 
\label{TC}. 
\eeq
For simple metals the theoretical critical temperature $T_c$, 
reported in the last column of Table 1,
is in good agreement with the experimental one $T_c^{exp}$: 
the relative difference $(T_c^{exp}-T_c)/T_c^{exp}$ is 
not large (i.e. within $10\%$), and for some metals 
(Tl, In, Sn) it is quite small (i.e. within $2\%$). 

Taking into account Eq. (\ref{fraction}), 
the BCS theory predicts that the zero-temperature condensate 
density in superconductors in the weak-coupling regime, can be written as 
\beq 
n_0(0) = 1.39 \, N(0) k_B T_c \; .  
\label{pip}
\eeq 

This equation resembles the familiar BCS result 
$\Delta(0) = 1.764 \, k_B T_c$ for the zero-temperature energy gap. 

We stress that the predictions of 
the BCS theory can be surely improved 
by using the Eliashberg theory \cite{carbotte,eliashberg}. 
This more sophisticated approach will not change the order of magnitude 
of the numbers in the first two columns of Tab. 1 
but it could change  the last significant figure.  

Coming back to the study of finite-temperature effects, 
in the upper panel of Fig. 1 
we plot the condensate density $n_0(T)$ vs electron-phonon 
strength $gN(0)$ for different values of the temperature $T$. 
As expected, by increasing 
the temperature $T$ it is necessary to increase the strength $gN(0)$ 
to get the same condensate density. 

\begin{figure}
\begin{center}
{\includegraphics[height=4.5in,clip]{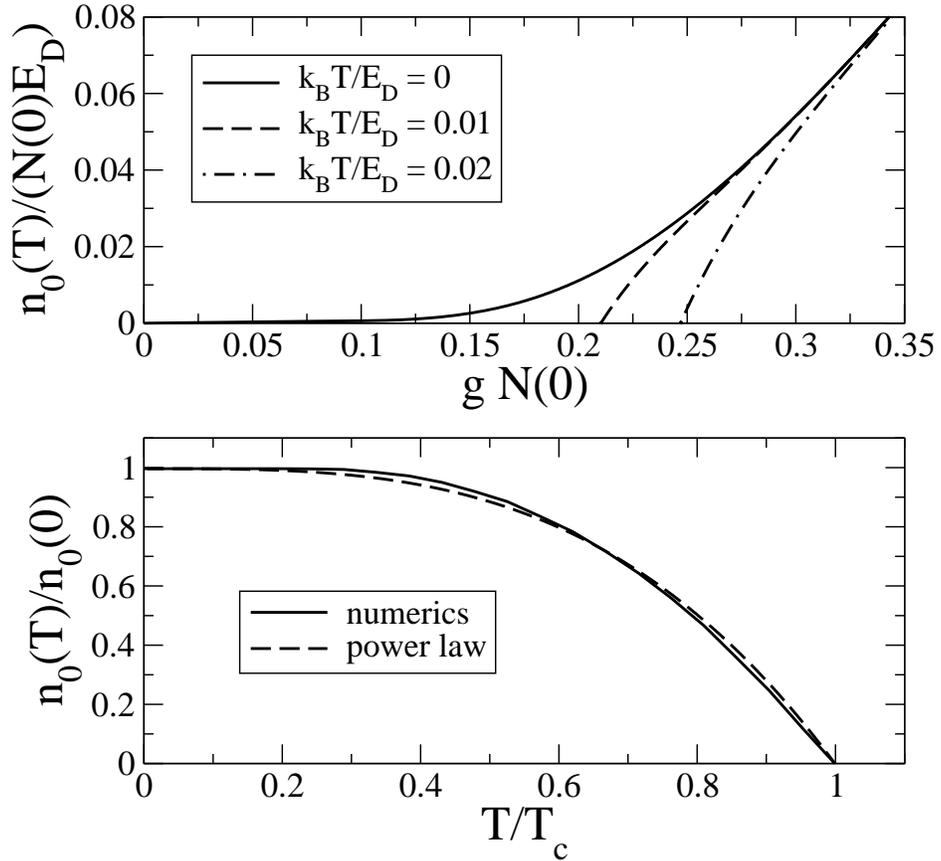}}
\end{center}
\caption{Upper panel: condensate density $n_0(T)$ 
vs electron-phonon strength $gN(0)$ in a superconductor 
for different values of the temperature $T$, 
where $N(0)$ is the density of states and $E_D=\hbar\omega_D$ 
is the Debye energy. 
Lower panel: condensate density $n_0(T)$ as a function of 
the temperature $T$ in a superconductor with 
$gN(0)=0.2$. Solid line: numerical solution 
of Eqs. (\ref{gap-1}) and (\ref{cond-1}); 
dashed line: analytical approximation, Eq. (\ref{approx}).}
\label{Fig1}
\end{figure}

As it happens for the energy gap $\Delta (T)$, one may show that 
Eqs. (\ref{cond-1}) and (\ref{cond-1}) together  also imply that
the condensate density may be written as its value at $T=0$ 
times a universal function of $T/T_c$. 
In the lower panel of Fig. 1 we plot the condensate density $n_0(T)/n_0(0)$ 
as a function of the temperature $T/T_c$: in the full range of temperatures
the numerical results (solid line) are reasonably well approximated 
by (dashed line) 
\beq 
n_0(T) = n_0(0) \ \left( 1 - \Big({T\over T_c}\Big)^{\alpha} \right) \; , 
\label{approx} 
\eeq
with $\alpha=3.16$ (best fit).

\section{Superconductors vs ultracold atoms} 

In metallic superconductors there is an attractive interaction 
between fermions only near the Fermi surface \cite{leggett1,fetter}. 
On the contrary, as remarked in the introduction, 
in superfluid ultracold two-component Fermi atoms, 
the effective inter-atomic interaction can be made 
attractive for all atoms of the system by using the technique 
of Fano-Feshbach resonances \cite{zwierlein2,leggett1,stringa}. 
This implies that in the BCS equations for ultracold atoms 
there is not a natural ultraviolet cutoff. 
For attractive ultracold atoms the mean-field BCS theory is given by the 
gap equation (\ref{gap}) and the number equation 
\beq 
N = \sum_{\bf k} \big( v_k^2 + 2(u_k^2-v_k^2) \bar{n}_k \big) \; ,     
\label{number} 
\eeq
while the condensate fraction is given by Eq. (\ref{cond}). 
But, for ultracold atoms, in these equations 
the sum over momenta is no more restricted within a thin shell around 
the Fermi surface. As well known, due to the 
choice of a contact potential, the gap equation (\ref{gap}) 
diverges in the ultraviolet. This divergence is logarithmic in 
two dimensions (2D) and linear in three dimensions (3D). 

In 3D, a suitable regularization is obtained by introducing 
the inter-atomic scattering length $a_F$ via the equation 
\beq
{m \over 4 \pi \hbar^2 a_F} = - {1 \over g} + 
{1 \over V} \sum_{\bf k} \frac1{2\epsilon_k} \,,
\eeq 
where $\epsilon_k=\hbar^2k^2/(2m)$ with $m$ the atomic mass, 
and then subtracting this equation from 
the gap equation \cite{eagles,leggett,noziers}. 
In this way one obtains the 3D regularized gap equation 
\beq 
-{m \over 4 \pi \hbar^2 a_F} = {1 \over V} 
\sum_{\bf k} \left( { \tanh{(\beta E_k/2)} \over 2 E_k} 
- \frac1{2\epsilon_k} 
\right) . 
\label{bcs3d}  
\eeq 
In 2D, quite generally 
the bound-state energy $\epsilon_B$ exists for any value 
of the interaction strength $g$ between two atoms 
\cite{marini,randeria}. For the contact potential 
the bound-state equation is 
\beq 
{1 \over g} = 
{1 \over V} \sum_{\bf k} \frac{1}
{{\hbar^2k^2\over 2m} + \epsilon_B} \, ,
\eeq 
and subtracting this equation from 
the gap equation \cite{marini,randeria} 
one obtains a 2D regularized gap equation 
\beq 
\sum_{\bf k} \left( 
\frac{1}{ {\hbar^2k^2\over 2m} + \epsilon_B} 
- { \tanh{(\beta E_k/2)} \over 2 E_k} \right) = 0 \; . 
\label{bcs2d}  
\eeq 
The number equation (\ref{number}) and the renormalized gap equation 
(\ref{bcs3d}) (or Eq. (\ref{bcs2d}) in 2D) are the so-called 
generalized BCS equations, from which one determines, for a fixed 
value of the temperature $T$ and the average number of atoms $N$, 
the chemical potential $\mu(T)$ and the gap energy $\Delta(T)$ 
as a function of the scattering length $a_F$ (or of the 
bound-state energy $\epsilon_B$ in 2D). The extended 
BCS equations can be applied in the full crossover from weak coupling 
to strong-coupling \cite{leggett1,stringa}. 
In 3D, the crossover is from the BCS state 
of weakly-interacting Cooper pairs (with $1/a_F \ll -1$) 
to the Bose-Einstein Condensate (BEC) of molecular dimers 
(with $1/a_F \gg 1$) across the unitarity 
limit ($1/a_F=0$) \cite{leggett}. In 2D, there is a similar 
BCS-BEC crossover by increasing the value $\epsilon_B$ 
of the bound-state energy \cite{marini,sala-odlro2}. 

At zero-temperature, by using the continuum limit 
$\sum_{\bf k} \to V/(2\pi)^3 \int d^3{\bf k}
\to V/(2\pi^2) \int k^2 dk$, the 3D condensate density 
(\ref{cond}) has a simple analytical expression \cite{sala-odlro}. 
The 3D density of states is $N(\xi)=(2m/\hbar^2)^{3/2}
\sqrt{\xi + \mu}/(4\pi^2)$ and the 3D condensate density 
is given by 
\beq 
n_0(0) = {m^{3/2} \over 8 \pi \hbar^3} \,
\Delta(0)^{3/2} \sqrt{{\mu(0)\over \Delta(0)}+
\sqrt{1+{\mu(0)^2 \over \Delta(0)^2} }} 
\; .   
\eeq 
In the 3D BCS regime ($1/a_F \ll -1$), 
where $\mu(0)/\Delta(0) \gg 1$ 
and the size of weaklybound Cooper pairs exceeds 
the typical interparticle spacing $k_{\rm F}^{-1}$, 
$\mu(0)$ approaches the non-interacting Fermi energy 
$\epsilon_F = \hbar^2 k_F^2/(2m)$ with $k_F=(3\pi^2 n)^{1/3}$ 
and there is an exponentially small energy gap $\Delta(0)= 8 e^{-2} 
\epsilon_F \exp{(\pi/(2k_F a_F))}$. In this weak-coupling regime 
the 3D condensate density becomes \cite{sala-odlro}  
\beq
n_0(0) = {1\over \pi} N(0) \Delta(0) 
= \frac{3\pi}{2 e^2}\, n\, \exp\!\left(\frac{\pi}{2k_Fa_F}\right)\, . 
\eeq
Notice that this is formula is similar to Eq. (\ref{e2}) 
of weak-coupling superconductors (here $a_F<0$), but 
the behavior of $\Delta(0)$ is quite different. 

In 2D, the density of states is constant and 
reads $N(\xi)=N(0)=(2m/\hbar^2)/(4\pi)$. 
The zero-temperature 
2D condensate density is easily obtained\cite{sala-odlro2} as 
\beq 
n_0(0) = {1\over 4} N(0) \Delta(0) 
\left( {\pi\over 2} + \arctan{\Big({\mu(0)\over \Delta(0)}\Big)} \right) \; ,  
\eeq 
while the zero-temperature 2D energy gap is given by 
the implicit formula 
\beq 
\Delta(0) = 2 \epsilon_F \Big(
\sqrt{1+{\mu(0)\over \Delta(0)}} - {\mu(0)\over \Delta(0)} \Big) \; . 
\eeq
From these equations, in the 2D BCS regime 
($0\le \epsilon_B \ll \epsilon_F$) where $\mu(0)/\Delta(0) \gg 1$ 
one finds exactly Eq. (\ref{e2}), but here the energy gap $\Delta(0)$ 
depends on the Fermi energy $\epsilon_F$ and the bound-state energy 
$\epsilon_B$ according to the formula \cite{randeria}  
\beq 
\Delta(0) = \sqrt{2 \epsilon_F \epsilon_B} \; , 
\eeq
while the chemical potential is $\mu(0)=\epsilon_F-{\epsilon_B}/2$. 
It is not surprising that in the BCS regime 
the condensate density of 2D 
superfluid atoms is formally equivalent 
to the Eq. (\ref{e2}) we have found for weak-coupling 
superconductors.  
In fact, to obtain Eq. (\ref{e2}) we have used the approximation 
$\int N(\xi) d\xi \simeq N(0) \int d\xi$ that is exact in the 
strictly 2D case, and the condition $\Delta(0)\ll \hbar \omega_D$ 
which implies that the upper limit of integration 
is practically $+\infty$. 

In the previous section we have shown that the BCS equations can 
be used to determine the (quite small) condensate fraction 
of superconductors only in the weak-coupling regime. 
Instead, the extended BCS equations have been 
used in recent papers \cite{sala-odlro,ortiz,ohashi,sala-odlro2} 
to get the condensate fraction of ultracold atoms 
in the full BCS-BEC crossover. The theory predicts that 
in the crossover the zero-temperature condensate fraction 
grows from zero to one. Two experiments 
\cite{zwierlein,inada} have confirmed these predictions 
for the 3D superfluid two-component Fermi gas. 

\section{Conclusions}

In this paper we have studied, within the mean-field BCS theory 
of superconductors, the condensate of electronic Cooper pairs 
at zero and finite temperature 
showing the crucial role played by the Debye frequency and by
the electron-phonon interaction. We have found that the 
zero-temperature condensate fraction $f(0)$ 
of weak-coupling metals is quite small ($\simeq 10^{-5}$) 
and the condensate density increases in metals with higher 
critical temperature $T_c$, 
according to the law $f(0)= 1.39\, N(0) k_BT_c$, 
where $N(0)$ is the density of states at the Fermi energy. 
As discussed by Chakravarty and Kee \cite{sudip}, 
the spin-spin correlation function depends significantly on the 
condensate density and magnetic neutron scattering can 
provide a direct measurement of the condensate fraction 
of a superconductor. In the next future our BCS predictions, 
which are meaningful for weak-coupling superconductors, 
could be experimentally tested. In the last part of the paper 
we have shown similarities and differences between metallic 
superconductors and atomic Fermi vapors in the determination 
of the condensate fraction by using the mean-field BCS theory 
and its extension in the BCS-BEC crossover.  

The author thanks A.J. Leggett and F. Toigo for useful comments 
and critical remarks.

\end{document}